\documentclass[twoside]{article}
\usepackage{fleqn,espcrc2}

\newcommand{\AmS}{{\protect\the\textfont2
  A\kern-.1667em\lower.5ex\hbox{M}\kern-.125emS}}

\def\bea{\begin{eqnarray}}
\def\eea{\end{eqnarray}}
\def\be{\begin{equation}}
\def\ee{\end{equation}}

\def\barr{\begin{eqnarray}}
\def\earr{\end{eqnarray}}

\newcommand{\ec}[1]{(\ref{#1})}

\def\beq{\begin{equation}}
\def\eeq{\end{equation}}
\def\barr{\begin{eqnarray}}
\def\earr{\end{eqnarray}}

\def\a{\alpha}
\def\b{\beta}

\def\d{\delta}

\def\e{\epsilon}

\def\m\mu
\def\n{\nu}

\def\s{\sigma}

\def\x{\xi}

\def\p{\partial}

\title{D-branes, Symplectomorphisms\\and Noncommutative Gauge Theories}

\author{I. Martin$^1$, J. Ovalle $^4$
and A. Restuccia$^2$ $^3$ \\ {\small Departamento de F\'{\i}sica, Universidad Sim\'on
Bol\'{\i}var, Venezuela} \\ {\small and} \\ {\small $^1$
Theoretical Physics Group, Imperial College , London University.
e-mail: isbeliam@usb.ve; isbeliam@ic.ac.uk}
\\ {\small $^2$ Department of Mathematics, King's College, London
University. e-mail: arestu@usb.ve}  \\ {\small $^3$ Invited talk at SSQFT,
Kharkov 2000.} \\ {\small $^4$ e-mail: jovalle@usb.ve }}
\begin{document}
\begin{abstract}
It is shown that the dual of the double compactified D=11
Supermembrane and a suitable compactified D=10 Super 4D-brane with
nontrivial wrapping on the target space may be formulated as
noncommutative gauge theories. The Poisson bracket over the
world-volume is intrinsically defined in terms of the minima of
the hamiltonian of the theory, which may be expressed in terms of
a non degenerate 2-form. A deformation of the Poisson bracket in
terms of the Moyal brackets is then performed. A noncommutative
gauge theory in terms of the Moyal star bracket is obtained. It is
shown that all these theories may be described in terms of
symplectic connections on symplectic fibrations. The world volume
being its base manifold and the (sub)group of volume preserving
diffeomorphisms generate the symplectomorphisms
which preserve the (infinite dimensional) Poisson bracket of the
fibration.
 \vspace{1pc}
\end{abstract}

\maketitle

\section{Introduction}
The formulation of D-brane theories in the presence of constant
antisymmetric background fields and its relation to
noncommutative gauge theories has recently attracted a lot of
interest \cite{con5}-\cite{giro}. It may well be that a
noncommutative formulation of the D=11 Supermembrane and the
Super M5-brane may indeed improve the understanding of the quantum
aspects of these theories. The spectrum of the D=11 Supermembrane
on a Minkowski target space was shown to be continuous from zero
to infinite \cite{dwln5}. However not much it is known about the
spectrum of the theory when the target space is compactified
\cite{russo}-\cite{mrt5}. Even less it is known about the spectrum of
the M5-brane. Nevertheless, one may extrapolate some known aspects from
the Supermembrane case since both theories are U-dual. The covariant
formulation of the M5-brane was found in \cite{pst} and
\cite{bln}. However the analysis of its physical hamiltonian, the
existence of singular configurations, the topological instabilities
and related problems have not yet been discussed in a conclusive
way. It is possible that a formulation of these theories in terms
of a noncommutative geometry may allow an improvement in their
analysis.

We describe in this talk a general approach to reach that
formulation. The first step is to introduce a symplectic geometry
intrinsic to the theory. This may be done when the target space is
suitably compactified. The non degenerate closed 2-form associated
to the symplectic geometry may be obtained in a general way from
the analysis of the Born-Infield action. We discuss this problem
in section 2. The second step in the construction is to obtain the
hamiltonian of the D-brane with non trivial wrapping on the target
space. We perform this construction for the double compactified
D=11 Supermembrane \cite{mor5}, \cite{mor2} and the compactified 4
D-brane in 10 dimensions. It turns out that the minima of these
hamiltonians are described by the dual configurations
introduced in \cite{br} and  in section 2. The final step
is to introduce the geometrical objects describing the
noncommutative formulation. This is done in terms of a symplectic
fibration and a symplectic connection over it. We also consider
deformations of the brackets introduced in these theories,
allowing a construction of noncommutative gauge theories in terms
of the usual Moyal star product. There is a precise one to one
correspondence in the sense of Kontsevich, between the original
theories and their deformations.
\section{The  dual configurations}
The Born-Infeld theory formulated over a Riemannian manifold $M$
may be described by the following $D$ dimensional action
\begin{equation}
S\left(A\right)=\int\limits_{M}\left(\sqrt{\det\left(g_{ab}+bF_{ab}\right)}
-\sqrt{g}\right)d^{D}x \label{Ac1} \end{equation} where $g_{ab}$
is an external euclidean metric over the compact closed manifold
$M$. $F_{ab}$ are the components of the curvature of connection
1-form $A$ over a $U(1)$ principle bundle on $M$.$b$ is a constant
parameter.

We may express the $\det\left(g_{ab}+bF_{ab}\right)$, using the
general formula obtained in \cite{br}, as \barr
\det\left(g_{ab}+bF_{ab}\right) &=& g\sum\limits_{m=0}^{n}
a_{m}b^{2m}{\ast\left[{\bf P}_{m}\wedge {\ast {\bf P}_{m}}\right]} \nonumber \\
&\equiv& gW, \earr where
\begin{equation}
{\bf P}_{m} \equiv\underbrace{F\wedge \ldots \wedge F}_{m}
\end{equation}
$a_{m}$ are known constants, see \cite{br}.

The first variation of (\ref{Ac1}) is given by \begin{equation}
\delta S\left(A\right)=\int\limits_{M}
W^{-\frac{1}{2}}\sum\limits_{m} m a_{m} b^{2m} d\delta A\wedge
{\bf P}_{m-1}\wedge {\ast {\bf P}_{m}} \end{equation} which yields the following
field equations
\begin{equation}
\sum\limits_{m} m a_{m} b^{2m}
{\bf P}_{m-1} \wedge d\left(W^{-\frac{1}{2}} {\ast {\bf P}_{m}}\right) =0.
\label{Ec. de Campo} \end{equation}

We introduce now a set $\mathcal{A}$ of $U(1)$ connection 1-forms
over $M$ \cite{br}. They are defined by the following conditions,
\begin{equation} {\ast  {\bf P}_{m}} \left(A\right) = k_{m}
{\bf P}_{n-m}\left(A\right) \hspace{1cm} ,m=0,\ldots,n, \label{dual}
\end{equation} where $n=\frac{D}{2}$,i.e we assume the
dimension $D$ of $M$ to be an even natural number. (\ref{dual})
is the condition that the Hodge dual transformation maps the
set $\left\{{\bf P}_{m}\hspace{1mm}, \hspace{3mm} m=0,\ldots,n\right\}$
into itself.

We observe that these connections, if they exist in a $U(1)$
principle bundle over $M$, are solutions of the field equations
(\ref{Ec. de Campo}).

In fact, (\ref{dual}) implies
\begin{equation}
\ast [{\bf P}_{m}\wedge \ast {\bf P}_{m}]=k_{m} \ast
[{\bf P}_{m}\wedge
{\bf P}_{n-m}]=k_{m} \ast {\bf P}_{n}
\end {equation}
but from (\ref{dual}), for $m=n$, we obtain
\begin{equation}
\ast {\bf P}_{n}=k_{n}
\end{equation}
which is constant. We thus have, for these
connections,
\begin{equation} W=\mathrm{constant}. \end{equation}

Finally, it results \begin{equation}
d\left(W^{-\frac{1}{2}}{\ast {\bf P}_{m}}\right)=k_{m}
W^{-\frac{1}{2}}d\left({\bf P}_{n-m}\right)=0, \end{equation} showing
that (\ref{dual}), if they exits, define a set of solutions to the
Born-Infeld field equations.

Let us analyse a particular case of (\ref{dual}). Let us consider
$n=\frac{D}{2}=1$. We then have \begin{equation} {\ast
{\bf P}_{1}}={\ast F} =
k_{1}. \label{mono1} \end{equation}

This solution represents a monopole connection over the $D=2$
manifold $M$. When $M$ is the sphere $S_{2}$, \ec{mono1} defines
the $U(1)$ connection describing the Dirac monopole on the Hopf
fibring $S_{3}\rightarrow S_{2}$. The constant $k_{1}$ is
determined from the condition \begin{equation} \int\limits_{M}
F=2\pi\times\mathrm{integer}\end{equation} which is a necessary
condition to be satisfied for a $U(1)$ connection, $F$ being its
curvature.

This solution was extended to $U(1)$ connections over Riemann
surfaces of any genus in \cite{mr5} . In \cite {mor5}  it was shown that they
describe the minima of the hamiltonian of the double compactified
$D=11$ supermembrane dual.

\section{Hamiltonian formulation}
The hamiltonian formulation of the double compactified D=11
Supermembrane dual was obtained in \cite{mor2}. Its hamiltonian
density in the light cone gauge is the following
\bea {\cal H}=&&\hspace{-0.7cm}\frac{1}{2}\frac{1}{\sqrt{W}}\left(
P^{M}P_{M}+det(\p_aX^M\p_bX_M)\right.\nonumber\\&&\hspace{-1.1cm}+(\Pi^a_r\p_aX^M)^2+
\frac{1}{4}(\Pi^a_r\Pi^b_s\e_{ab}\e^{rs})^2
 \left.+\frac{1}{4}W(\ast F^r)^2\right)\nonumber\\&&\hspace{-1.2cm}
-A^r_0\p_c\Pi^c_r+\Lambda\e^{ab}\p_b\left(\frac{\p_aX^MP_M+\Pi^c_rF^r_{ac}}{\sqrt{W}}\right)
\label{Hamil4} \eea
where $P_M$ are the conjugate momentum to $X^M$ while $\Pi^a_r$
are the corresponding momentum to $A^r_a$. The index $r$ denote
the 2 compactified directions on the target space. $a$ is the
world volume index while $M$ label the LCG transverse directions
in the target space.

Its supersymmetric extension may be obtained in an straightforward
way from the supermembrane hamiltonian in the LCG by the procedure
described in \cite{mor2}.

We may solve explicitly the constraints on $\Pi^c_r$ obtaining \be
\Pi^c_r=\e^{cb}\p_b\Pi_r;\hspace{1.2cm}r=1,2
 \ee
Defining the 2-form $\omega$ in terms of $\Pi_r$ as

 \be
\omega  = \p_a \Pi_r \p_b \Pi_s\e^{rs} d\xi^{a} \wedge  d\xi^{b}, \ee the
condition of non trivial membrane winding imposes a restriction on
it,  namely
\begin{equation}
\oint_{\Sigma } \omega =2 \pi n .
\end{equation}

With this condition on $\omega $, Weil's theorem ensures that
there always exist an associated $U(1)$ principal bundle over
$\Sigma $ and a connection on it such that $\omega $ is its
curvature. The minimal configurations for the hamiltonian
(\ref{Hamil4}) may be expressed in terms of such connections.

In \cite{mor5} the minimal configurations of the hamiltonian of
the double compactified supermembrane were obtained. In spite of
the fact that the explicit expression (\ref{Hamil4}) was not then
obtained, all the minimal configurations were found. They
correspond to $\Pi_{r}$ = $\hat{\Pi}_{r}$ satisfying \be \ast
\hat{\omega} =
\e^{ab}\p_a\hat{\Pi}_r\p_b\hat{\Pi}_s\e^{rs}=n\sqrt{W}\hspace{1cm}n\neq0
\ee

The explicit expressions for $\hat{\Pi}_r$ were obtained in that
paper \cite{mor5}. As mentioned before, they correspond to $U(1)$
connections on non trivial principle bundles over $\Sigma$. The
principle bundle is characterized by the integer $n$ corresponding
to an irreducible winding of the supermembrane. Moreover the
semiclassical approximation of the hamiltonian density around the
minimal configuration, was shown to agree with the hamiltonian
density of super Maxwell theory on the world sheet, minimally
coupled to the seven scalar fields representing the coordinates
transverse to the world volume of the super-brane.

As mention in the introduction these minima correspond to the
 dual solutions of section 2 corresponding to 2D-brane. We
now consider the hamiltonian of the D=10 4D-brane. It may be
obtained by the following double dimensional reduction procedure.
We start from the PST action for the super M5-brane. We consider
the gauge fixing condition which fixes the scalar field to be
proportional to the world volume time. We then perform the usual
double dimensional reduction by taking one of the target space
coordinates $X^{11}=\sigma^5$ where $\sigma^5$ is one of the world
volume local coordinates. After several calculations we end up
with the following canonical lagrangian
\begin{equation}
{\cal L}= P_m\dot{X}^m+P^{ij}\dot{B}_{ij}-{\cal H}_c
\end{equation}
\begin{equation}
{\cal H}_c = \lambda\phi+\lambda^i\phi_i+\theta_i\partial_jP^{ij}
\end{equation}
where \bea
\phi=&&\hspace{-0.6cm}\frac{1}{2}P^2+2g+2\left(\frac{1}{8}P^{ij}P^{kl}g_{ik}g_{jl}+
{\ast }H^{i} \hspace{-0.1cm}{\ast }H^{j}g_{ij}\right)\nonumber\\
&&\hspace{-0.7cm}
+\frac{1}{32}\left(\frac{1}{4}\epsilon_{ijkl}P^{ij}P^{kl}\right)^2
\eea
\begin{equation}
\phi_l=P_m\partial_lX^m+\frac{1}{2}\epsilon_{ijkl}P^{ij}{\ast }H^k
\end{equation}
\begin{equation}
{\ast }H^i=\frac{1}{6}\epsilon^{ijkl}H_{jkl}
\end{equation}

We finally obtain the hamiltonian in the LCG

\bea {\cal H}=\hspace{-0.7cm}&&\frac{1}{\sqrt{W}}\left(
\frac{1}{2}P^{M}P_{M}+2g+2\left(\frac{1}{8}P^{ij}P^{kl}g_{ik}g_{jl}\right.\right.\nonumber\\
&& \left. + {\ast }H^{i} \hspace{-0.1cm}{\ast }H^{j}g_{ij}\right)+\frac{1}{32}\left(\frac{1}{4}\epsilon_{ijkl}P^{ij}P^{kl}\right)^2\nonumber\\
&&+\Lambda^{lq}\partial_q\left(P_M\partial_lX^M+\frac{1}{2}\epsilon_{ijkl}P^{ij}{\ast }H^k\right)
\label{Hamil76} \eea

where $\Lambda^{lq}$ are antisymmetric lagrange multipliers
associated to the generator of volume preserving diffeomorphisms.

There is also a global constraint given by
\begin{equation}
\oint_{C_i}\left(P_M\partial_lX^M+\frac{1}{2}\epsilon_{ijkl}P^{ij}{\ast }H^k\right)d\sigma^l=0
\end{equation}
where $C_i$ is a basis of homology of dimension 1. We will not
consider any further this global constraint. We will work only
with the diffeomorphisms connected to the identity.

We now consider a target space with 4 compactified directions,
$M_6$x$S^1$x$S^1$x$S^1$\-x$S^1$. We construct the dual formulation
associated to the compactified directions. We associate to each
$X_r$, $r=1,...,4$ a $B^3$  3-form
\begin{equation}
dX_r\rightarrow dB^3_r
\end{equation}
It is more convenient to work with the Hodge dual
of the 3-form,
\begin{equation}
B_{rijk}\rightarrow A^l_r
\end{equation}
in the spatial world volume sector $B_{rijo}$ are
Lagrange multipliers.
We denote $\Pi_{rl}$ the conjugate momenta to $A^l_r$.

There is a constraint on $\Pi_{rl}$ which yields
\begin{equation}
\Pi_{rl}=\partial_l\Pi_r
\end{equation}
We may then perform a canonical transformation to obtain
\begin{equation}
\Pi_{rl}\dot{A}^l_r=\dot{\Pi}_r(\partial_lA^l_r)
\end{equation}
that is, $\partial_lA^l_r$ is the conjugate momenta to $\Pi_r$:
\bea
 \Pi_r\equiv{\cal A}_r\nonumber\\
 \partial_lA^l_r\equiv{\it \Pi}_r
\eea We then obtain the dual formulation to (\ref{Hamil76}).

For the determinant of the induced metric we obtain \bea
&&\frac{1}{4!}\left(\epsilon^{i_1...i_4}\partial_{i_1}X^{a_1}...\partial_{i_4}X^{a_4}\right)^2
\nonumber\\
&&\rightarrow\frac{1}{4!}\left(\epsilon^{i_1...i_4}\partial_{i_1}X^{b_1}...
\partial_{i_4}X^{b_4}\right)^2
\nonumber\\
&&+\frac{1}{3!}\left(\epsilon^{i_1...i_4}\partial_{i_1}{\cal
 A}_r\partial_{i_2}X^{b_2}...\partial_{i_4}X^{b_4}\right)^2
\nonumber\\
&&+\frac{1}{2!}\left(\epsilon^{i_1...i_4}\partial_{i_1}{\cal
A}_r\partial_{i_2}{\cal
 A}_s\partial_{i_3}X^{b_3}\partial_{i_4}X^{b_4}\right)^2
\nonumber\\
&&+\left(\epsilon^{i_1...i_4}\partial_{i_1}{\cal
A}_r\partial_{i_2}{\cal A}_s\partial_{i_3}{\cal
A}_t\partial_{i_4}X^{b_4}\right)^2
\nonumber\\
&&+\left(\epsilon^{i_1...i_4}\partial_{i_1}{\cal
A}_r\partial_{i_2}{\cal A}_s\partial_{i_3}{\cal
A}_t\partial_{i_4}{\cal A}_u\right)^2 \eea where the index $b$ is
used to denote the non compactified directions.

For the terms quadratic on the momenta to the antisymmetric field
$B_{ij}$ we obtain similar terms, the connection 1-forms ${\cal
A}_r$ replaces the corresponding terms where the compactified
coordinates appear. In the same way the momenta ${\it \Pi}^r$
replaces the conjugate momenta of the compactified coordinates.
That is, in the dual formulation (with the additional canonical
transformation we mentioned above) the compactified coordinates
are replaced by the connection 1-forms ${\cal A}_r$. However, it is
clear from the dual formulation that ${\cal A}_r$ may have non
trivial transitions of a very specific form over a nontrivial
bundle. This fact is difficult to realize in terms of the original
maps from the world volume to the compactified directions of the
target space.
\section{The noncommutative formulation}
We now introduce a symplectic 2-form in the previous formulation.
We take
\begin{equation}
F_{ij}d\sigma^i\wedge d\sigma^j
\end{equation}
where $F_{ij}$ is the curvature of the connection 1-form which
minimize the hamiltonian of the theory. For the D=11 Supermembrane
we obtained
\begin{equation}
\ast F=n,\hspace{2.5cm}i,j=1,2
\end{equation}
as discussed before.

For the D=10 Super 4D-brane we obtain
\begin{equation}
F\propto \ast F
\end{equation}
\begin{equation}
\ast (F\wedge F)\propto  n
\end{equation}
over the 4 dimensional spatial world volume. These are two of the
dual solutions introduced in \cite{br} and explained
in section 2.

The procedure to obtain the symplectic noncommutative
formulation for the double compactified D=11 Supermembrane was
explicitly introduced in \cite{mor2}. We will now obtain a similar
formulation for the 4 D-brane in 10 dimensions.

The approach uses the symplectic 2-forms previously introduced to
obtain a description of the theory in terms of  symplectic
connections over a symplectic fibration.

We consider the metric $W^{ij}$ defined by
\begin{equation}
W^{ij}\equiv\hat{\Pi}^i_r\hat{\Pi}^j_r
\end{equation}
where
\begin{equation}
\hat{\Pi}^i_r\equiv F^{ij}\partial_j \hat{\cal A}_r\label{g54}
\end{equation}
The metric $W^{ij}$ is taken to be the metric over the spatial
world volume for which $F^{ij}$ satisfies the duality conditions. $\hat{\Pi}^i_r$ is a well defined vielbein.
(\ref{g54}) defines $\hat{\cal A}_r$.

We then introduce the following Poisson bracket over the world
volume
\begin{equation}
\{B,C\}\equiv F^{ij}\partial_iB\partial_jC\label{pos}
\end{equation}
It satisfies the Jacobi identity

\begin{eqnarray}
\{\{A,B\},C\}+\{\{C,A\},B\}+\{\{B,C\},A\}&& \nonumber \\
=\left(F^{kl}F^{ij}+F^{jl}F^{ki}+F^{il}F^{jk}\right)
&&\nonumber \\
.D_l\left(\p_iA\p_jB\p_kC\right)\nonumber\\=k\frac{\epsilon^{klij}}{\sqrt{W}}D_l(\p_iA\p_jB\p_kC)=0
\end{eqnarray}
Where $D_l$ denotes the covariant derivative with respect to the
metric $W^{ij}$. $A$, $B$ and $C$ are scalar fields.

We now introduce the rotated covariant derivative
\begin{equation}
D_r\equiv\hat{\Pi}^i_{r}D_i
\end{equation}
\begin{equation}
{\cal D}_r=D_r+\{{\cal A}_r, \}
\end{equation}
and curvature
\begin{equation}
{\cal F}_{rs}=D_r{\cal A}_s-D_s{\cal A}_r+\{{\cal A}_r,{\cal
A}_s\}
\end{equation}

The hamiltonian density of the double compactified Supermembrane
was expressed in terms of these geometrical objects in
\cite{mor2}. We now perform the analogous formulation for the 4
D-brane.

We obtain
\begin{eqnarray}
&&\frac{1}{4!}\left[\epsilon^{i_1...i_4}\p_{i_1}X^{b_1}...\p_{i_4}X^{b_4}
\right]^2\nonumber \\
&&\rightarrow
\frac{1}{4!}\left[\{X^{b_1},X^{b_2}\}\{X^{b_3},X^{b_4}\} \right]^2
\end{eqnarray}
\begin{eqnarray}
&&\frac{1}{3!}\left[\epsilon^{i_1...i_4}\p_{i_1}{\cal A
}_r\p_{i_2}X^{b_2}...\p_{i_4}X^{b_4} \right]^2\nonumber
\\ &&\rightarrow\frac{1}{3!}\left[{\cal D}_rX^{b_2}.\{X^{b_3},X^{b_4}\}
\right]^2
\end{eqnarray}
\begin{eqnarray}
&&\frac{1}{2!}\left[\epsilon^{i_1...i_4}\p_{i_1}{\cal A
}_r\p_{i_2}{\cal A }_s\p_{i_3}X^{b_3}\p_{i_4}X^{b_4}
\right]^2\nonumber \\ &&\rightarrow \frac{1}{2!}\left[{\cal
F}_{rs}.\{X^{b_3},X^{b_4}\} \right]^2
\end{eqnarray}
\begin{eqnarray}
&&\left[\epsilon^{i_1...i_4}\p_{i_1}{\cal A }_r\p_{i_2}{\cal A
}_s\p_{i_3}{\cal A }_t\p_{i_4}X^{b_4} \right]^2\nonumber \\ &&
\rightarrow \left[{\cal F}_{rs}{\cal D}_tX^b\} \right]^2
\end{eqnarray}
\begin{eqnarray}
\left[\epsilon^{i_1...i_4}\p_{i_1}{\cal A }_r...\p_{i_4}{\cal A}_u
\right]^2\rightarrow \left[{\cal F}_{rs}{\cal F}_{tu}\right]^2
\end{eqnarray}
where it is understood that the antisymmetric part on the $b$
indexes and the target space $r$, $s$,... indexes is taken.

Similar formulae may be written for all other terms in the
hamiltonian density for the 4 D-brane. They may be rewritten in
terms of the bracket (\ref{pos}), the covariant derivative and the
curvature ${\cal F}_{rs}$.

The expression for the double compactified D=11 Supermembrane was
\cite{mor2}: \bea &&H=\int_\Sigma {\cal
H}=\int_\Sigma\frac{1}{2\sqrt{W}}\left[(P^M)^2+({\it
\Pi^r})^2\right.\nonumber\\
&&+\frac{1}{2}W\{X^M,X^N\}^2+W({\cal D}_rX^M)^2\nonumber
\\
&&\left.+\frac{1}{2}W({\cal F}_{rs})^2 \right]+
\int_\Sigma\left[\frac{1}{8}\sqrt{W}n^2\right.\nonumber\\
&&\left.-\Lambda\left({\cal D}_r{\it
\Pi}^r+\{X^M,P_M\}\right)\right]\hspace{1cm}\label{Hamil5} \eea

The geometrical interpretation of the above formulation
(\ref{Hamil5}) was given in \cite{mor2} in terms of symplectic
fibrations and  connection 1-form over it. We have shown here that
the same geometrical description may be given for the 4 D-brane.
We will discuss it in more detail shortly. Before then, we would
like to remark that there is a natural deformation of the above
formulations in terms of the Moyal bracket. Let us see how this
deformation may be realized preserving the symplectic structure on
the fibration, for the case of the double compactified
Supermembrane.

We replace in (\ref{Hamil5}) the Poisson bracket
\begin{equation}
\{B,C\}=F^{ij}\p_iB\p_jC\label{h6}
\end{equation}
by the Moyal bracket $\{,\}_M$. (\ref{h6}) is the first term in
the expansion of the Moyal bracket. We notice that under the gauge
transformation generated by the first class constraint
\begin{equation}
\d X^M=\{\xi,X^M\}
\end{equation}
\begin{equation}
\d {\cal A}_r=-{\cal D}_r\xi=-\left(D_r\xi+\{{\cal
A}_r,\xi\}_M\right)
\end{equation}
\begin{equation}
\d {\cal D}_rX^M=\{\xi,{\cal D}_rX^M\}_M
\end{equation}

These properties ensure that the hamiltonian density transform as
\begin{equation}
\d {\cal H}=\{\xi,{\cal H}\}_M
\end{equation}
The integral over a compact world volume renders the canonical
lagrangian invariant under the gauge transformations.

We notice that the physical degrees of freedom of both theories,
the double compactified D=11 Supermembrane (\ref{Hamil5}) and its
Moyal deformation, are exactly the same. Moreover it can be shown
that there is a one to one correspondence, in the sense of
Kontsevich, between both theories.

In \cite{mor2} a geometrical description of the symplectic non
commutative gauge theory was introduced. The same geometrical
interpretation may be used to describe its deformation in terms of
the Moyal bracket and the compactified 4 D-brane we have
discussed.

We consider a symplectic fibration with base manifold the spatial
world volume, which is a closed (without boundary) manifold. Over
the fibration we consider the (infinite dimensional) Poisson
bracket of the sections $X^M(\s)$, $P_M(\s)$:
\begin{equation}
[X^M(\s),P_M(\s')]=\d(\s,\s')
\end{equation}
This Poisson structure is preserved under the transition maps of
the fibration. These maps are defined by
\begin{equation}
\d X^M=\{\x,X^M\}
\end{equation}
\begin{equation}
\d P_M=\{\x,P_M\}
\end{equation}
over $U_\a \bigcap U_\b$ , ${U_\a}$ is a covering of the base
manifold. We notice that the transformation maps are defined in
terms of the (finite dimensional) Poisson bracket or Moyal bracket
over the world volume, and they preserve the (infinite
dimensional) Poisson bracket on the fibration. ${\cal A}_r$ define
a symplectic connection over the fibration. That is, the Poisson
bracket on the fibration is preserved under the holonomy generated
by ${\cal A}_r$.

The three theories we have discussed, the double compactified D=11
Supermembrane, its Moyal deformation and the compactified 4
D-brane all admit the same geometrical interpretation. They
describe the dynamics of a symplectic connection over a symplectic
fibration.
\section{Conclusions}
We formulated the double compactified D=11 Supermembrane and the
compactified Super 4 D-brane in terms of a symplectic noncommutative gauge theory. We constructed a deformation of the
compactified D=11 Supermembrane in terms of the Moyal brackets.
There exist a one to one correspondence
between both theories in the sense of Kontsevich. A unified geometrical description of these
theories was given in terms of a symplectic fibration over the
world volume and the dynamics of symplectic connections over it. We
hope the analysis for the D=10 4D-brane may be extended to
describe the M5-brane in 11 dimension. Once that formulation is
available we may start analysing the corresponding quantum field
theory. We hope to report on that shortly.

\end{document}